\documentclass[prl,10pt,a4paper,twocolumn,notitlepage,showpacs,superscriptaddress]{revtex4-1}

\usepackage{amsmath}
\usepackage{times}
\usepackage{amssymb}
\usepackage[dvipdfm]{graphicx}
\usepackage[unicode=true, pdfusetitle,
 bookmarks=true,bookmarksnumbered=false,bookmarksopen=false,
 breaklinks=false,pdfborder={0 0 1},backref=false,colorlinks=true,dvipdfm]
 {hyperref}
\newcommand{\be}{\begin{equation}}
\newcommand{\ee}{\end{equation}}
\newcommand{\ben}{\begin{eqnarray}}
\newcommand{\een}{\end{eqnarray}}
\newcommand{\bes}{\begin{subequations}}
\newcommand{\ees}{\end{subequations}}

\begin{document}

\title{Unidimensional reduction of the 3D Gross-Pitaevskii equation with two- and three-body interactions}

\author{W. B. Cardoso}
\author{A. T. Avelar}
\affiliation{Instituto de F\'isica, Universidade Federal de Goi\'as, 74.001-970, Goi\^ania, Goi\'as, Brazil.}
\author{D. Bazeia}
\affiliation{Departamento de F\'isica, Universidade Federal da Para\'iba, 58051-970, Jo\~ao Pessoa, Para\'iba, Brazil.}

\begin{abstract}
We deal with the three-dimensional Gross-Pitaevskii equation, which is used to describe a cloud of dilute bosonic atoms that interact under competing two- and three-body scattering potentials.
We study the case where the cloud of atoms is strongly confined in two spatial dimensions, allowing us to build an unidimensional nonlinear equation, controlled by the nonlinearities and the confining potentials that trap the system along the longitudinal coordinate. We focus attention on specific limits, dictated by the cubic and quintic coefficients, and we implement numerical simulations to help us to quantify the validity of the procedure.
\end{abstract}
\pacs{03.75.Lm, 05.45.Yv}

\maketitle

\section{Introduction}

A cloud of dilute bosonic atoms weakly interacting confined in magnetic traps and cooled down to extremely low temperatures ($\sim \mu$K) can behave as a Bose-Einstein
condensate (BEC), as firstly demonstrated in vapors of rubidium \cite{AndersonSCi95} and sodium \cite{DavisPRL95}. The existence of BECs \cite{Pethick02,Pitaevskii03} has triggered a lot of new investigations, with a diversity of scenarios being proposed and tested \cite{BurgerPRL99,DenschlagSCi00,StreckerNat02,KhaykovichSCi02,Kivshar03,DalfovoRMP99,LeggettRMP01,Malomed06}.
The presence of experimental techniques for manipulating the strength of the effective interaction between trapped atoms \cite{RobertsPRL01} leads us to believe that in BECs we have an excellent opportunity to investigate localized solutions of atomic matter waves taking advantage of Feshbach-resonance management \cite{InouyeNat98,TimmermansPR99,KevrekidisPRL03}.

The case of BECs with weak interactions is standard, and the atomic distribution is well described by the three-dimensional (3D) Gross-Pitaevskii equation (GPE) \cite{Pethick02,Pitaevskii03}. However, if the confinement in the transversal directions is stronger than the longitudinal one, the dynamics is then governed by the one-dimensional (1D) GPE \cite{Abdullaev04}. The 1D reduction of the 3D GPE using a Gaussian variational approach \cite{PGPRA98,JacksonPRA98}, assuming a stronger confinement in the transversal directions, has shown that a nonpolynomial Schr\"odinger equation (NPSE) is the effective equation that describes the atomic distribution in the longitudinal direction \cite{SalasnichPRA02,SalasnichLP02}. In this sense, the cubic nonlinear Schr\"odinger equation (CNLSE) is obtained when the system evolves under the weak coupling regime, which consists in taking $g_3|\psi|^2<<1$, where $g_3$ represents the cubic nonlinearity of the system, and $|\psi|^2$ describes the atomic density in the BEC.

Other recent works have considered confinement patterns for several distinct geometries \cite{SalasnichPRA07,AdhikariNJP09,SalasnichPRA09,FilatrellaPRA09,SalasnichJPA09,GligoricCH09,YoungPRA10}. For instance, in \cite{SalasnichPRA07} the authors derived an effective NPSE for quasi-1D trap with periodic modulation, and in \cite{AdhikariNJP09} an effective nonlinear Schr\"odinger equation (NLSE) for cigar-shaped and disc-shaped Fermi superfluid was suggested. Moreover, in Ref. \cite{SalasnichPRA09} the authors considered solitons and solitary vortices in pancake-shaped BECs by reduction of the 3D GPE through a variational approach, and in Ref. \cite{SalasnichJPA09} a generalized NPSEs for matter waves under anisotropic transverse confinement was deduced. Application with Feshbach-resonance management to a tightly confined BEC was considered in \cite{FilatrellaPRA09}, and in Ref. \cite{YoungPRA10} the authors introduced dimensional reduction of a binary Bose-Einstein condensate with distinct dimensionality, e.g., with the confining potentials of the two components acting in a way such that each component belongs to different space dimensions.

All the papers cited above consider that the reduced effective equation is obtained from the 3D cubic GPE, but it is known that there are systems which engender cubic and quintic nonlinearities. For example, in BECs the presence of quintic nonlinearity is in general due to three-body interactions, and the cubic and quintic nonlinearities may be used to describe the scenario where the two-body scattering is somehow weakened in a way such that the three-body effects cannot be neglected anymore \cite{nature,RoatiPRL07,3B,3body}. Hence it is relevant to consider the 1D reduction of the more general case of 3D cubic-quintic GPE (CQGPE). In the present work we follow this route, and we study the 1D reduction of a 3D GPE which is used to describe a BEC with cubic and quintic nonlinearities, prepared to describe a cloud of dilute bosonic atoms weakly interacting with competing two- and three-body interactions. The main motivation comes from recent studies, in which we investigated 1D GPE with cubic and quintic nonlinearities \cite{AvelarPRE09,AvelarNARWA10}. Although the cubic nonlinearity is usually present in a cloud of bosonic atoms weakly interacting and cooled down to extremely low temperatures to form a BEC, there are systems which also requires the quintic nonlinearity.

In the standard procedure, the 1D CQGPE is obtained via expansions of the cubic nonlinearity, and the quintic term is always negative \cite{SalasnichPRA02}. In the present case, we show how to get to the 1D CQGPE  where the quintic nonlinearity can be found with positive or negative real value. Since the quintic nonlinearity is related to three-body interaction, it is of current interest. For instance, we know that such effect was observed in a BEC of cesium atoms arising due to the Efimov resonance \cite{ChinPRL05}. Moreover, the three-body interaction becomes more significant for higher density and larger scattering length, as found in atomic chip and in atomic-wave guide \cite{CQ-mot}.

We follow the variational approach, in which we consider a strong confinement in the transverse $(y,z)$ directions. A generalized NPSE is then obtained to describe the 1D profile along the longitudinal $x$ direction. In the regime of weak coupling, this equation is shown to reduce to a cubic-quintic GPE. To see how this works, in the next section we present the model and study the derivation of the 1D equation. The work follows with Sec.~III, where we investigate some specific issues related to the 1D GPE. We end the paper in Sec.~IV, where we present comments and conclusions.

\section{Derivation of 1D nonpolynomial equation}

The Hamiltonian that describes a cloud of weakly interacting bosonic atoms, for which there are competing two- and three-body interactions can be written as follows \cite{Pethick02}
\begin{eqnarray}
\hat{H}&\!=\!& \int (\frac{\hbar ^2}{2m} \mathbf{\nabla} \hat{\Psi}^{\dagger} \mathbf{\nabla} \hat{\Psi} )d\textbf{r}\! +\! \frac{1}{2}\!\int \!\hat{\Psi}^{\dagger} \hat{\Psi}^{\prime\dagger} \mathcal{V}_2(\textbf{r},\textbf{r}^{\prime}) \hat{\Psi}^{\prime} \hat{\Psi} d\textbf{r} d\textbf{r}^{\prime} \nonumber \\ &+& \frac{1}{3}\int \hat{\Psi}^{\dagger} \hat{\Psi}^{\prime\dagger} \hat{\Psi}^{\prime\prime\dagger} \mathcal{V}_3(\textbf{r},\textbf{r}^{\prime},\textbf{r}^{\prime\prime}) \hat{\Psi}^{\prime\prime} \hat{\Psi}^{\prime} \hat{\Psi} d\textbf{r} d\textbf{r}^{\prime} d\textbf{r}^{\prime\prime},
\label{hamiltonian}
\end{eqnarray}
where $m$ is the atomic mass, $\hat{\Psi}^{\dagger}$ ($\hat{\Psi}$) is the field operator which creates (annihilates) an atom, $\textbf{r}=x\hat{i}+y\hat{j}+z\hat{k}$, $\nabla$ is the gradient, and $\mathcal{V}_2$ and $\mathcal{V}_3$ are the two- and three-body potentials, respectively.

When the system behaves as a BEC, its ground state is macroscopically populated allowing the use of the Bogoliubov approximation \cite{Pethick02}. In this case, one can replace the field operator in Eq.~(\ref{hamiltonian}) by a classical field $\psi(\textbf{r},t)$ which describes the density profile of the condensate. Also, assuming that the function $\psi(\textbf{r})$ varies slowly over distances of the order of the range of the interatomic forces, one can substitute $\textbf{r}^{\prime}$ and $\textbf{r}^{\prime\prime}$ for $\bf{r}$ in the argument of $\psi({\textbf{r}}^{\prime},t)$ and $\psi({\textbf{r}}^{\prime\prime},t)$, respectively. We use this together with the Heisenberg equation of motion to obtain the GPE. Its explicit form can be written as, using dimensionless quantities
\begin{equation}
i\psi_t = - \frac{1}{2}\nabla^2\psi + U\psi + 2\pi g_3|\psi|^2\psi + 3\pi^2g_5|\psi|^4 \psi.
\label{3dnlse}
\end{equation}
Here $\psi=\psi(\textbf{r},t)$ represents the wave-function of the BEC, $\nabla^2$ stands for the Laplacian, and $\partial\psi/\partial t= \psi_t$. Also, $U= U(\textbf{r},t)$ is an external potential produced by the magneto-optical trap, and $g_3= g_3(\textbf{r},t)$ and $g_5= g_5(\textbf{r},t)$ are the coefficients of the cubic and quintic nonlinearities, associated to the scattering of two and three particles. The positive or negative sign of $g_i, i=3,5,$ corresponds to repulsion or attraction between two or three atoms, respectively. The scattering length of the atoms in the dilute gas can be modulated by the Feshbach resonance, tuned via magnetic or optical external fields. 

We can consider a Lagrangian density, responsible to describe the 3D CQGPE (\ref{3dnlse}); it is given by
\begin{eqnarray}
\mathcal{L}&=&\frac{i}{2}\left( \psi^*\psi_t - \psi\psi_t^* \right) - \frac{1}{2}|\nabla\psi|^2 - \frac{1}{2}\left(y^2+z^2\right)|\psi|^2 \nonumber \\
&-& V(x)|\psi|^2 - \pi g_3|\psi|^4 - \pi^2g_5|\psi|^6,
\label{lagrangian}
\end{eqnarray}
where we are considering that the potential is split in two pieces, one, transversal, describing the transversal trapping, given by
\begin{equation}
V(y,z)=\frac{1}{2}\left(y^2+z^2\right),
\end{equation}
and the other being $V(x)$, which describes the longitudinal trapping, much weaker than the transversal one. 

Here we take into account the procedure shown in \cite{SalasnichPRA02}, and so we assume the following \textit{ansatz}:
\begin{equation}
\psi=\frac{1}{\sqrt{\pi}\sigma}\exp\left[-\frac{(y^2+z^2)}{2\sigma^2} \right]\phi,
\label{ansatz}
\end{equation}
where $\phi=\phi(x,t)$ and $\sigma=\sigma(x,t)$ are the longitudinal wave-function and the transverse width, respectively. An effective Lagrangian density is obtained through the substitution of (\ref{ansatz}) in (\ref{lagrangian}) and the integration over the transversal coordinates. In the process of doing the calculation, it is also necessary to use the adiabatic approximation \cite{SalasnichPRA02}, which consists of neglecting the spatial derivatives of the transverse width, supposing that it is constant, approximately. The calculations then follow easily, and we end up with an effective Lagrangian density which reads
\begin{eqnarray}
{\mathcal{L}}_{eff}&=&
\frac{i}{2} \big( \phi^{\ast }\frac{\partial \phi}{\partial t} 
- \phi\frac{\partial \phi^{\ast}}{\partial t} \big) 
- \frac{1}{2} \left\vert{\frac{\partial \phi}{\partial x}}\right\vert^{2} 
- \frac{1}{2\sigma^2} |\phi|^2
\nonumber 
\\
&-& \frac{\sigma^2}{2} |\phi|^2
- V(x,t)|\phi|^2 - \frac{1}{2} g_3 \frac{|\phi|^4}{\sigma^2} - \frac{1}{3} g_5 \frac{|\phi|^6}{\sigma^4}.
\label{effective}
\end{eqnarray}
Now, through variations of ${\mathcal{L}}_{eff}$ with respect to $\phi^{\ast}$ and $\sigma$ one finds two coupled equations:
\begin{eqnarray}
i\frac{\partial\phi}{\partial t} &=& -\frac{1}{2}\frac{\partial^2\phi}{\partial x^2} +  \left[ V(x) + \frac{1+s^2}{2s} \right]\phi \nonumber \\
&+& \frac{g_3}{s}|\phi|^2\phi + \frac{g_5}{s^2}|\phi|^4\phi,
\label{eff1}
\end{eqnarray}
\begin{eqnarray}
s^3-s(1+g_3|\phi|^2)-\frac{4}{3}g_5|\phi|^4=0,
\label{apr2}
\end{eqnarray}
where we are using $s=\sigma^2$, for simplicity.

The above Eq. (\ref{apr2}) can be solved via the Cardano's method. The formal solution can be written as 
\begin{eqnarray}
s=\frac{1}{3}\frac{ \left( 18g_5|\phi |^4 + 3\sqrt{A} \right)^{2/3} + 3(1+g_3|\phi |^2)  }
{ \left( 18g_5|\phi |^4 + 3\sqrt{A} \right)^{1/3} },
\label{ex1}
\end{eqnarray}
where
\begin{equation}
A= 36g_5^2|\phi |^8-3(1+g_3|\phi |^2)^3.
\end{equation} 
So, using (\ref{ex1}) we can rewrite (\ref{eff1}) in the form
\begin{equation}
i\frac{\partial\phi}{\partial t} = -\frac{1}{2}\frac{\partial^2\phi}{\partial x^2} +  V\phi + N_p \phi,
\label{np}
\end{equation}
where $N_p$ corresponds to the nonpolynomial term of the generalized GPE which is given by
\begin{equation}
N_p = \frac{ (3+\frac{C^2}{3B^2} )BC + 3B (g_3|\phi |^2C + 3g_5|\phi |^4B) }{2C^2},
\label{nonpa}
\end{equation}
where
\bes\label{nonpb}\ben B&=&(18g_5|\phi |^4 + 3\sqrt{A})^{1/3},
\\
C&=&B^2+3(1+g_3|\phi |^2).
\een\ees
This is the main result of this work, and it shows that the presence of the quintic term maintains the reduced 1D GPE nonpolynomial, but it changes significantly the nonpolynomial function which specifies the 1D effective equation. However, we note that when one sends the quintic nonlinearity to zero, that is, if we set $g_5=0$, the generalized Eq.~(\ref{np}) becomes
\be\label{cubic1}
i\frac{\partial\phi}{\partial t} = -\frac{1}{2}\frac{\partial^2\phi}{\partial x^2} +  V\phi + {\widetilde N}_p \phi,
\ee
where
\be\label{cubic2}
{\widetilde N}_p=\frac{1+(3/2)g_3|\phi|^2}{\sqrt{1+g_3|\phi|^2}},
\ee
which exactly reproduces the equation obtained in \cite{SalasnichPRA02}.

\begin{figure}
\includegraphics[width=5cm]{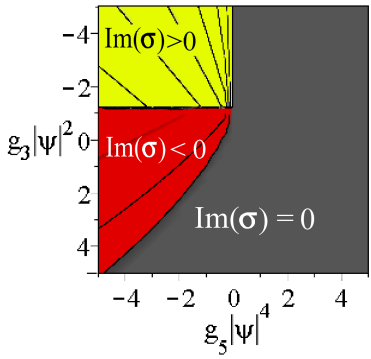}
\caption{(Color online) Imaginary values of the transverse width $\sigma$ versus $g_3|\phi |^2$ and $g_5|\phi |^4$. Note that in the regions Im$(\sigma)<0$ and Im$(\sigma)>0$ solutions are not allowed.}
\label{F0}
\end{figure}

Due to the fact that $\sigma$ has to be real and positive, some regions for $g_3|\phi |^2$ and $g_5|\phi |^4$ are not allowed as one can see in Fig. \ref{F0}, where we display the imaginary values of $\sigma$, in function of $g_3|\psi |^2$ and $g_5|\psi |^4$. Of course, the region with Im$(\sigma)=0$ is the region physically accessible for the {\textit {ansatz}} \eqref{ansatz}.

\section{Specific considerations}

With the general result obtained above, let us now consider some cases of current interest, in the weak coupling regime. We will focus attention on the three distinct cases: \textit{i}) $g_3|\phi|^2 \sim g_5|\phi|^4 \ll 1$, \textit{ii}) $g_5|\phi|^4 \ll g_3|\phi|^2 \ll 1$, and \textit{iii}) $g_3|\phi|^2 \ll g_5|\phi|^4 \ll 1$. In the case \textit{i}), we can expand the nonpolynomial contribution in Eq.~\eqref{ex1} in power series of $g_3|\phi|^2$ and $g_5|\phi|^4$ in order to get, up to first order,
\begin{equation}
\sigma\simeq 1+\frac{1}{4}g_3|\phi|^2+\frac{1}{3}g_5|\phi|^4,
\label{wc1}
\end{equation}
If we use (\ref{wc1}) in (\ref{eff1}), we obtain the familiar 1D CQGPE given by
\begin{equation}
i \frac{\partial \phi}{\partial t} = \Big[ - \frac{1}{2}
\frac{\partial^2}{\partial x^2} + U(x,t) + g_3 |\phi|^2 + g_5 |\phi|^4
\Big] \phi,
\label{1dcqnlse}
\end{equation}
where $U(x,t)=V(x,t)+1$. The Eq. (\ref{1dcqnlse}) describes the case of cubic and quintic nonlinearities, in the weak interacting regime. 
The other cases \textit{ii} and \textit{iii} can be obtained straightforwardly: the Eq. (\ref{1dcqnlse}) is reduced to its cubic or quintic form, replacing $g_5=0$ or $g_3=0$, respectively. The cubic 1D GPE was studied in Ref. \cite{SalasnichPRA02}. So, in the weak coupling regime, the 3D CQGPE reduces to a 1D CQGPE with the standard polynomial form. Recently, exact bright and dark soliton solutions for this equation with varying coefficients were considered in Ref. \cite{AvelarPRE09}.

Here we note that if we neglect the three-body interactions in the starting Hamiltonian \eqref{hamiltonian}, the procedure will lead us with the nonpolynomial
equation given by \eqref{cubic1}. From this equation, we can get to a polynomial equation with cubic and quintic interactions if one expands the nonpolynomial term \eqref{cubic2} up to second order in $g_3$. In this case, however, one gets to the quintic term with negative nonlinearity, and with the cubic coefficient much bigger then the quintic one. The case under investigation in the present work is more general, and we can get to negative or positive quintic nonlinearity, appropriate to describe attractive or repulsive quintic interactions.

\begin{figure}
\includegraphics[width=4cm]{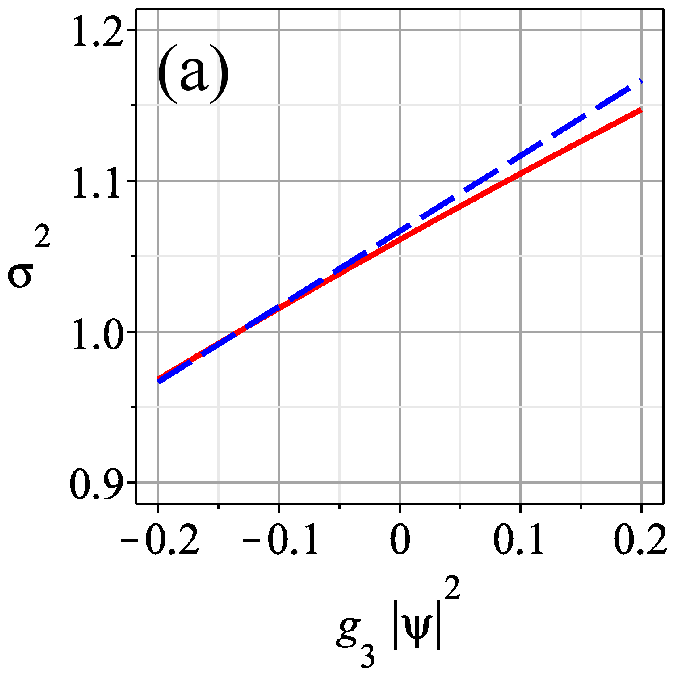}\hfil%
\includegraphics[width=4cm]{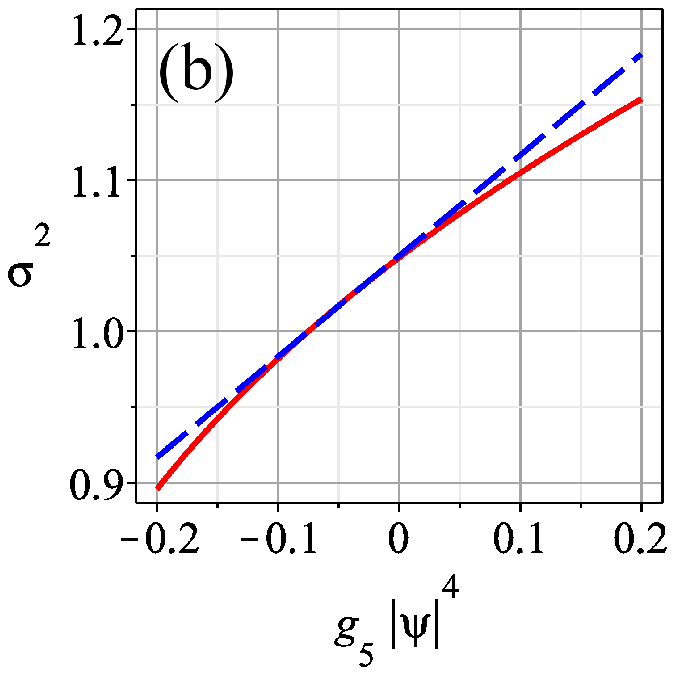}
\caption{(Color online) Plots of $\sigma^2$ given by Eqs.~(\ref{apr2}) and (\ref{wc1}) with solid (red) and dashed (blue) lines, respectively. In (a) and (b) we display $\sigma^2$ versus $g_3|\phi|^2$ (for $g_5|\phi|^4=0.1$) and versus $g_5|\phi|^4$ (for $g_3|\phi|^2=0.1$), respectively.}
\label{F1}
\end{figure}

\begin{figure}
\includegraphics[width=4cm]{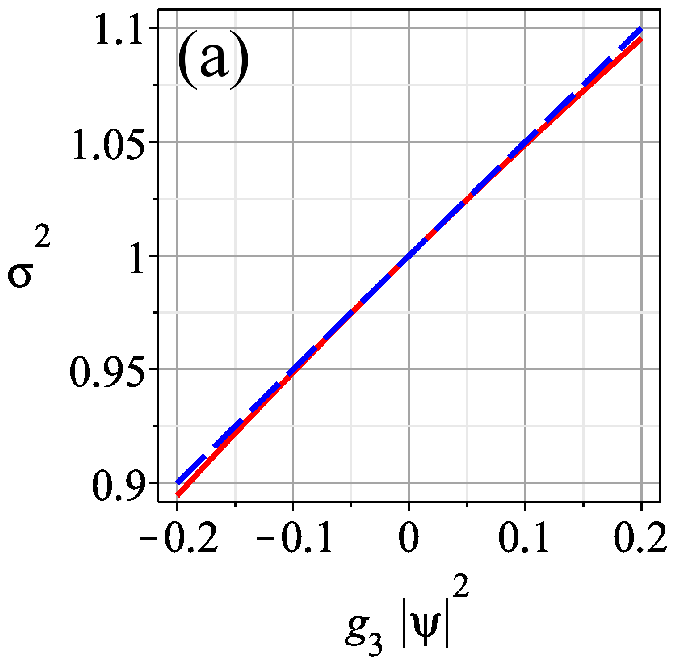}\hfil%
\includegraphics[width=4cm]{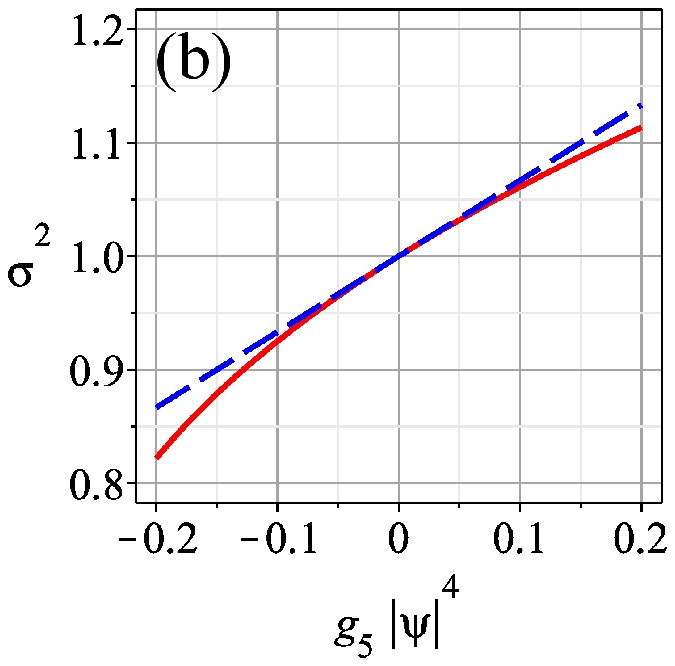}
\caption{(Color online) Plots of $\sigma^2$ given by Eqs.~(\ref{apr2}) and (\ref{wc1}), for the cases \textit{ii} and \textit{iii}, with solid (red) and dashed (blue) lines, respectively. In (a) and (b) we display $\sigma^2$ versus $g_3|\phi|^2$ (case \textit{ii}) and versus $g_5|\phi|^4$ (case \textit{iii}), respectively.}
\label{F2}
\end{figure}

To better understand the above procedure, let us investigate the square of the transverse width of the cigar-shaped condensate, $\sigma^2(x,t)$. In Figs. \ref{F1}a and \ref{F1}b we depict $\sigma^2$ in the appropriate region. The plots shown in these figures consider Eqs.~(\ref{apr2}) and (\ref{wc1}) versus $g_3|\phi|^2$ (with $g_5|\phi|^4=0.1$) and versus $g_5|\phi|^4$ (with $g_3|\phi|^2=0.1$), respectively. Also, in Figs.~\ref{F2}a and \ref{F2}b one depicts $\sigma^2$ for the cases \textit{ii} and \textit{iii}, respectively. The results suggest that the procedure is reliable.


\begin{figure}
\includegraphics[width=7cm]{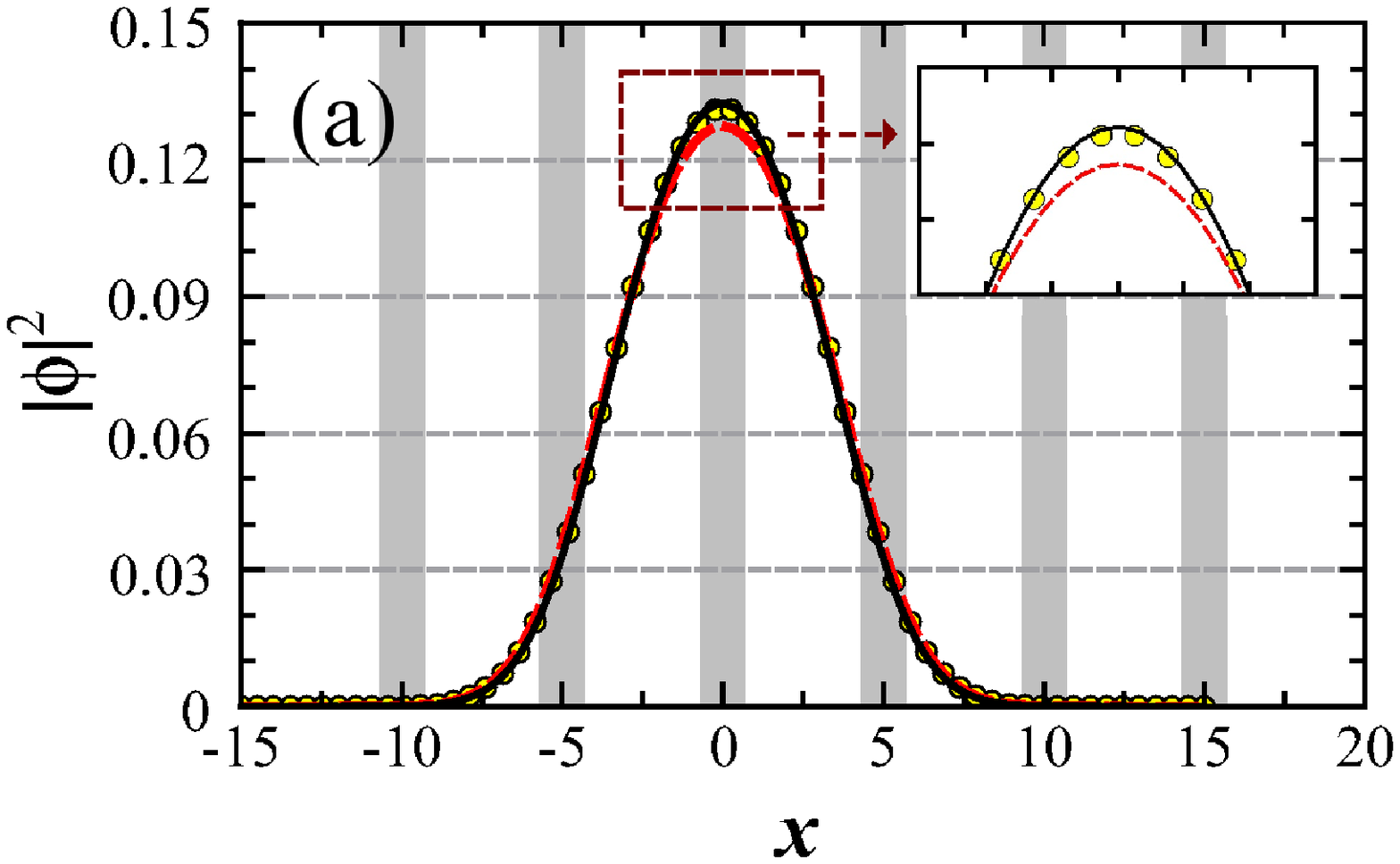}\hfil%
\includegraphics[width=7cm]{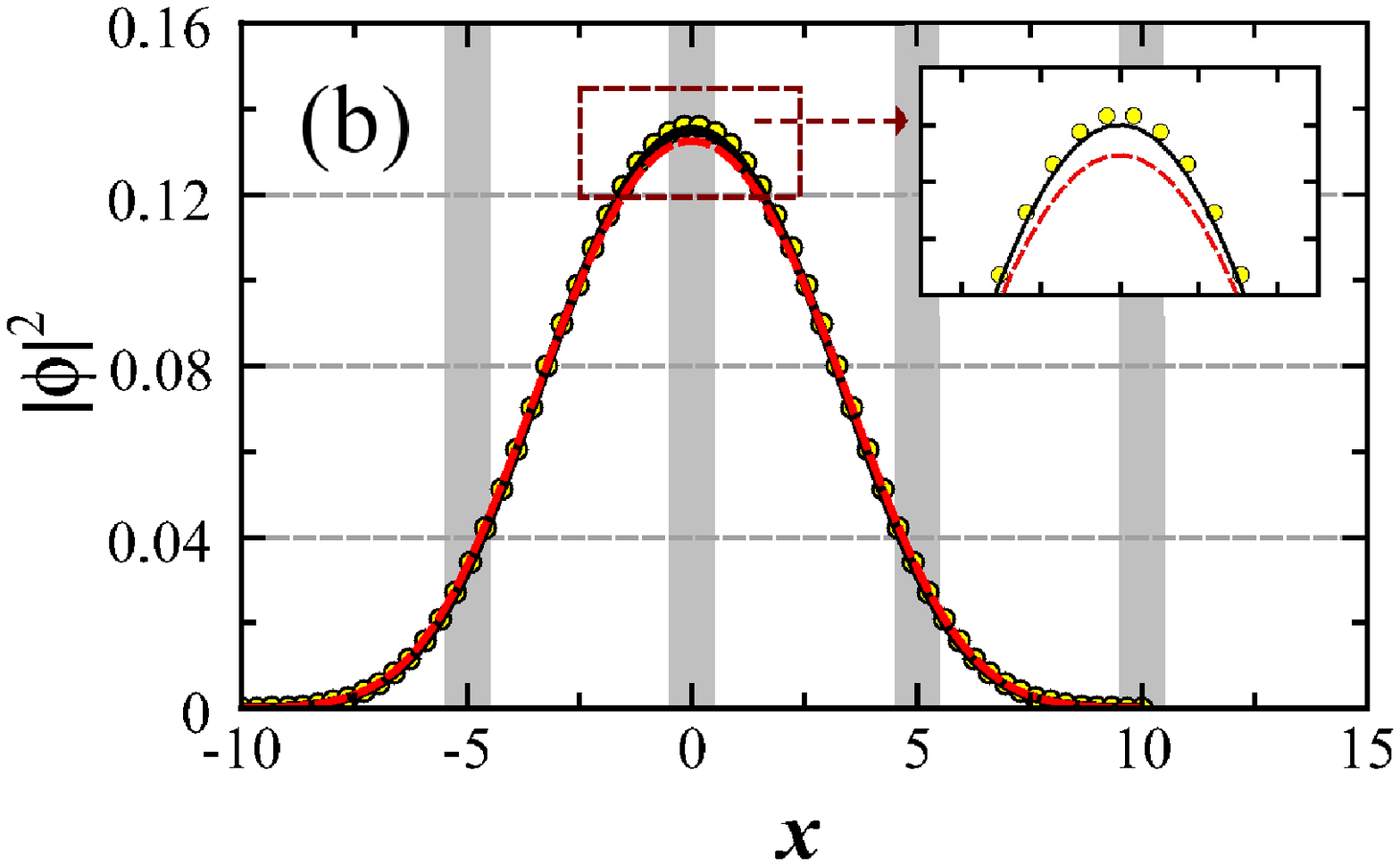}\hfil%
\includegraphics[width=7cm]{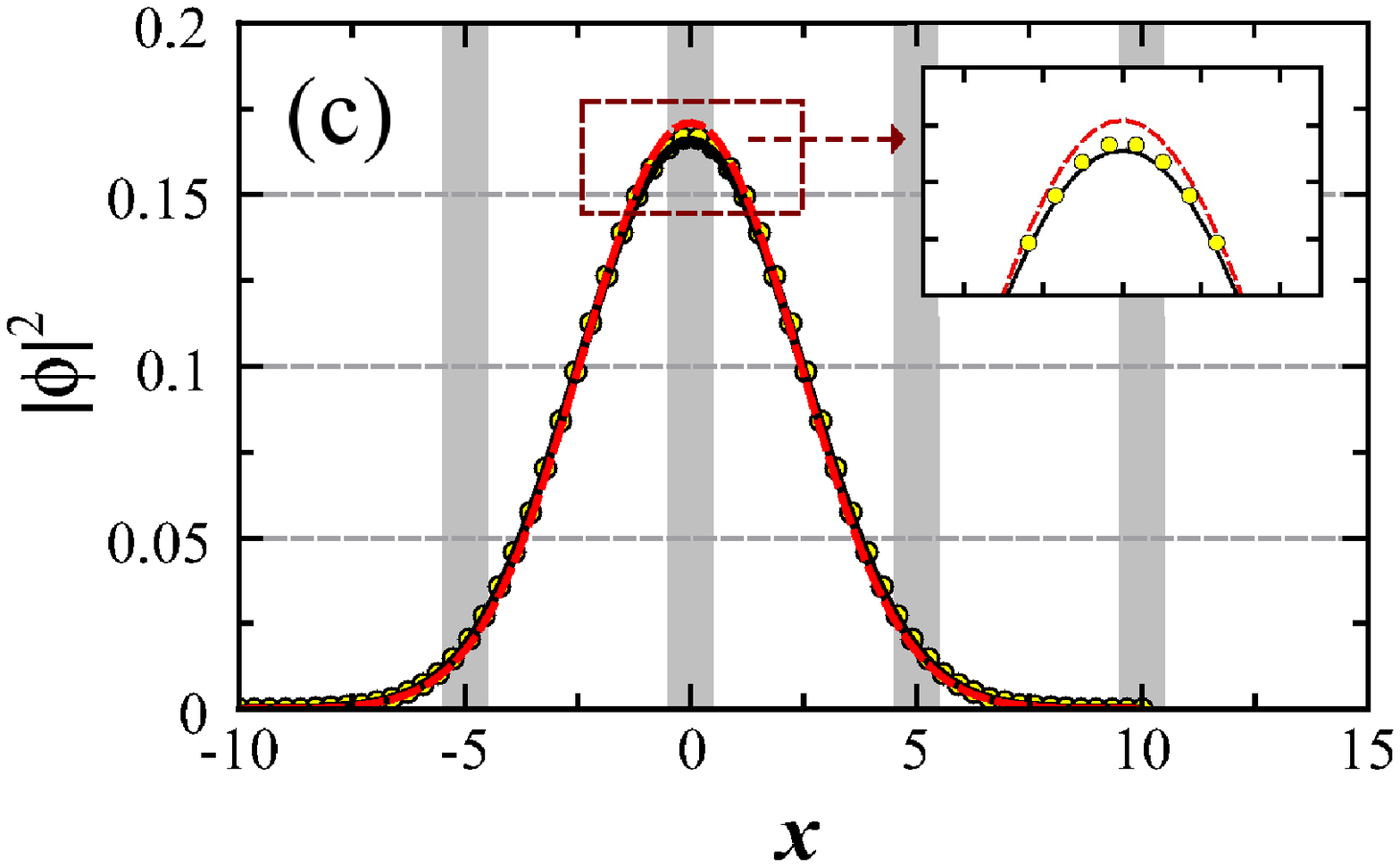}%
\caption{(Color online) Plots of the density profile $|\phi|^2$ considering a quadratic axial confinement, where $V(x)=(\lambda x)^2/2$, with $\lambda=0.1$. We depict the numerical solutions of the 3D GPE given by Eq.~(\ref{3dnlse}) with circles (yellow). We also depict numerical solutions of the generalized NPGPE given by Eq.~(\ref{np}) with the solid (black) curve, and of the 1D CQGPE given by Eq.~(\ref{1dcqnlse}) with the dashed (red) curve. For comparison, we display the cases: (a) $g_3=1.0$ and $g_5=1.0$, (b) $g_3=1.0$ and $g_5=0.0$ (cubic case), and (c) $g_3=0.0$ and $g_5=1.0$ (quintic case), respectively. The insets detail some specific regions shown in the figures. Note the complete agreement between the 3D GPE and the generalized NPGPE.}
\label{F3}
\end{figure}

We go further into the subject by implementing a detailed numerical investigation of both the 3D and 1D equations. In Fig. \ref{F3} we depict the profile of the solutions which propagates in imaginary time, using a numeric algorithm with \textit{split-step} based in the Crank-Nicolson method for the three Eqs.~(\ref{3dnlse}), (\ref{np}), and (\ref{1dcqnlse}) \cite{MuruganandamCPC09}. The solution of the 3D equation is projected in the $x$-axis in the form $\phi(x,t)=\int\int \textrm{d}y\textrm{d}z \psi(x,y,z,t)$, where $\psi(x,y,z,t)$ is the numerical solution of the 3D equation (\ref{3dnlse}). In this figure, we verify the complete agreement between the 3D GPE (\ref{3dnlse}), the generalized NPGPE (\ref{np}), and the 1D GPE (\ref{1dcqnlse}), which is here obtained as an approximation valid in the weak coupling regime. 

Now, comparing the nonpolynomial term of the generalized NPGPE [Eqs.~(\ref{nonpa}) and \eqref{nonpb}] with the corresponding polynomial term in the CQGPE ($1+g_3|\phi|^2+g_5|\phi|^4$), one gets that 
\begin{equation}
g_3=-\frac{4}{3}g_5 \mathrm{max}\{|\phi|^2\}.
\label{eq12}
\end{equation}
Thus, if the cubic and quintic nonlinearities satisfy the above equation, the generalized NPGPE (\ref{np}) leads to similar results when compared to those by the polynomial 1D CQGPE. On the other hand, when (\ref{eq12}) is far from being satisfied, we cannot ensure that the solutions of the 1D CQGPE describe the effective longitudinal profile of the atomic density in the condensate anymore. However, even in this case the NPGPE is shown to provide a valid description, when compared to the 3D GPE.

\section{Final comments}

In this work we introduced a direct investigation of 1D reduction of the 3D GPE with cubic and quintic nonlinearities under strong transversal confinement. The main result of this work is obtained in Eqs.~\eqref{np}, \eqref{nonpa} and \eqref{nonpb}, where one gets a generalized NPGPE that describes the longitudinal profile of a BEC with two- and three-body interactions.

We further used the 1D CQGPE to study the weak coupling limit, thus obtaining an effective polynomial equation that characterizes the atomic density when it is supposed to be described by the cubic and quintic nonlinearities in the weak coupling regime. The results are easily reduced to the case where the nonlinearity is purely cubic or purely quintic. 

A detailed numerical study was also implemented, to investigate both the 3D and the effective 1D equations, and some solutions were depicted in Fig.~\ref{F3}, to show the feasibility of the procedure which leads to the 1D effective equations.

\subsection*{Acknowledgments}

The authors would like to thank the Brazilian agencies CAPES, CNPq and FUNAPE/GO for partial financial support.


\begin{thebibliography}{99}

\bibitem{AndersonSCi95} M. H. Anderson, J. R. Ensher, M. R. Matthews, C. E. Wieman, and E. A. Cornell, Science \textbf{269}, 198 (1995).

\bibitem{DavisPRL95} K. B. Davis, M.-O. Mewes, M. R. Andrews, N. J. van Druten, D. S. Durfee, D. M. Kurn, and W. Ketterle, Phys. Rev. Lett. \textbf{75}, 3969 (1995).

\bibitem{Pethick02}C. J. Pethick and H. Smith, \textit{Bose-Einstein Condensation in Dilute Gases} (Cambridge University Press, Cambridge, 2002).

\bibitem{Pitaevskii03}L. Pitaevskii and S. Stringari, \textit{Bose-Einstein Condensation} (Oxford University Press, Oxford, 2003).

\bibitem{BurgerPRL99} S. Burger, K. Bongs, S. Dettmer, W. Ertmer, K. Sengstock, A. Sanpera, G. V. Shlyapnikov, and M. Lewenstein, Phys. Rev. Lett. \textbf{83}, 5198 (1999).

\bibitem{DenschlagSCi00} J. Denschlag, J. E. Simsarian, D. L. Feder, Charles W. Clark, L. A. Collins, J. Cubizolles, L. Deng, E. W. Hagley, K. Helmerson, W. P. Reinhardt, S. L. Rolston, B. I. Schneider, and W. D. Phillips, Science \textbf{287}, 97 (2000).

\bibitem{StreckerNat02} K. E. Strecker, G. B. Partridge, A. G. Truscott, and R. G. Hulet, Nature \textbf{417}, 150 (2002).

\bibitem{KhaykovichSCi02} L. Khaykovich, F. Schreck, G. Ferrari, T. Bourdel, J. Cubizolles, L. D. Carr, Y. Castin, and C. Salomon, Science 296, 1290 (2002).

\bibitem{Kivshar03} Y. S. Kivshar, G. P. Agrawal, \textit{Optical Solitons: From Fibers to Photonic Crystals} (Academic Press, San Diego, 2003).

\bibitem{DalfovoRMP99} F. Dalfovo, S. Giorgini, L. P. Pitaevskii, and S. Stringari, Rev. Mod. Phys. \textbf{71}, 463 (1999).

\bibitem{LeggettRMP01} A. J. Leggett, Rev. Mod. Phys. \textbf{73}, 307 (2001).

\bibitem{Malomed06} B. A. Malomed, \textit{Soliton Management in Periodic Systems} (Springer, New York, 2006).

\bibitem{RobertsPRL01} J. L. Roberts, N. R. Claussen, S. L. Cornish, E. A. Donley, E. A. Cornell, and C. E. Wieman, Phys. Rev. Lett. \textbf{86}, 4211 (2001).

\bibitem{InouyeNat98} S. Inouye, M. R. Andrews, J. Stenger, H.-J. Miesner, D. M. Stamper-Kurn, and W. Ketterle, Nature \textbf{392}, 151 (1998).

\bibitem{TimmermansPR99} E. Timmermans, P. Tommasini, M. Hussein, and A. Kerman, Phys. Rep. \textbf{315}, 199 (1999).

\bibitem{KevrekidisPRL03} P. G. Kevrekidis, G. Theocharis, D. J. Frantzeskakis, and B. A. Malomed, Phys. Rev. Lett. \textbf{90}, 230401 (2003).

\bibitem{Abdullaev04} F. Kh. Abdullaev and V. V. Konotop, \textit{Nonlinear Waves: Classical and Quantum Aspects} (Kluwer Academic Publishers, New York, 2004), p. 251.

\bibitem{PGPRA98} V. M. P\'{e}rez-Garc\'{i}a, H. Michinel, and H. Herrero, Phys. Rev. A \textbf{57}, 3837 (1998).

\bibitem{JacksonPRA98} A. D. Jackson, G. M. Kavoulakis, and C. J. Pethick, Phys. Rev. A \textbf{58}, 2417 (1998).

\bibitem{SalasnichPRA02} L. Salasnich, A. Parola, and L. Reatto, Phys. Rev. A \textbf{65}, 043614 (2002).

\bibitem{SalasnichLP02} L. Salasnich, Laser Phys. \textbf{12}, 198 (2002).

\bibitem{SalasnichPRA07} L. Salasnich, A. Cetoli, B. A. Malomed, F. Toigo, and L. Reatto, Phys. Rev. A \textbf{76}, 013623 (2007).

\bibitem{AdhikariNJP09} S. K. Adhikari and L. Salasnich, New J. Phys. \textbf{11}, 023011 (2009).

\bibitem{SalasnichPRA09} L. Salasnich and B. A. Malomed, Phys. Rev. A \textbf{79}, 053620 (2009).

\bibitem{SalasnichJPA09} L. Salasnich, J. Phys. A: Math. Theor. \textbf{42}, 335205 (2009).

\bibitem{FilatrellaPRA09} G. Filatrella, B. A. Malomed, and L. Salasnich, Phys. Rev. A \textbf{79}, 045602 (2009).

\bibitem{GligoricCH09} G. Gligori\'c, A. Maluckov, L. Salasnich, B. A. Malomed, and Lj. Had\v{z}ievski, CHAOS \textbf{19}, 043105 (2009).

\bibitem{YoungPRA10} L. E. Young-S., L. Salasnich, and S. K. Adhikari, Phys. Rev. A \textbf{82}, 053601 (2010).

\bibitem{nature}T. Kraemer, M. Mark, P. Waldburger, J. G. Danzl, C. Chin, B. Engeser, A. D. Lange,
K. Pilch, A. Jaakkola, H.-C. N\"agerl, and R. Grimm, Nature \textbf{440}, 315 (2006).

\bibitem{RoatiPRL07}G. Roati, M. Zaccanti, C. D'Errico, J. Catani, M. Modugno, A. Simoni, M. Inguscio, and G. Modugno, Phys. Rev. Lett. \textbf{99}, 010403 (2007).

\bibitem{3B}L. Mazza, M. Rizzi, M. Lewenstein, and J.I. Cirac, Phys. Rev. A \textbf{82}, 043629 (2010).

\bibitem{3body}Y. Wang, J. P. D'Incao, and B.D. Esry, \textit{Ultracold three-body collisions near narrow Feshbach resonances,} [arXiv:0906.5019].

\bibitem{AvelarPRE09} A. T. Avelar, D. Bazeia, and W. B. Cardoso, Phys. Rev. E \textbf{79}, 025602(R) (2009).

\bibitem{AvelarNARWA10}W. B. Cardoso, A. T. Avelar, and D. Bazeia, Nonlinear Anal.: Real World Appl. \textbf{11}, 4269 (2010).

\bibitem{ChinPRL05} C. Chin, T. Kraemer, M. Mark, J. Herbig, P. Waldburger, H.-C. N\"agerl, and R. Grimm, Phys. Rev. Lett. \textbf{94}, 123201 (2005).

\bibitem{CQ-mot} F. Kh. Abdullaev, A. Gammal, L. Tomio, and T. Frederico, Phys. Rev. A \textbf{63}, 043604 (2001); A. Bulgac, Phys. Rev. Lett. \textbf{89}, 050402 (2002); A. E. Leanhardt, A. P. Chikkatur, D. Kielpinski, Y. Shin, T. L. Gustavson, W. Ketterle, and D. E. Pritchard, Phys. Rev. Lett. \textbf{89}, 040401 (2002); P. Pieri and G. C. Strinati, Phys. Rev. Lett. \textbf{91}, 030401 (2003); T. K\"ohler, Phys. Rev. Lett. \textbf{89}, 210404 (2002); Ai-Xia Zhang and Ju-Kui Xue, Phys. Rev. A \textbf{75}, 013624 (2007); U. Roy, R. Atre, C. Sudheesh, C. N. Kumar, and P. K. Panigrahi, J. Phys. B: At. Mol. Opt. Phys. \textbf{43}, 025003 (2010).

\bibitem{MuruganandamCPC09} P. Muruganandam and S. K. Adhikari, Comput. Phys. Commun. \textbf{180}, 1888 (2009).

\end{thebibliography}
\end{document}